# Spectrally multiplexed heralded single photon source at telecom-band

Hao Yu[1,6,*], Chenzhi Yuan[1,*], Ruiming Zhang[1], Zichang Zhang[1], Hao Li[5], You Wang[1,3], Guangwei Deng[1], Lixing You[5], Haizhi Song[1,3], Zhiming Wang[1,7], Guang-Can Guo[1,4], and Qiang Zhou[1,2,4]

[1]*Institute of Fundamental and Frontier Sciences, University of Electronic Science and Technology of China, Chengdu 610054, China.*
[2]*Yangtze Delta Region Institute (Huzhou) & School of Optoelectronic Science and Engineering, University of Electronic Science and Technology of China, Huzhou 313001, China.*
[3]*Southwest Institute of Technical Physics, Chengdu 610041, China.*
[4]*CAS Key Laboratory of Quantum Information, University of Science and Technology of China, Hefei 230026, China.*
[5]*Shanghai Institute of Microsystem and information Technology, Chinese Academy of Sciences, Shanghai 200050, China.*
[6]*Institut national de la recherche scientifique - Centre Énergie, Matériaux et Télécommunications (INRS-EMT), Varennes J3X 1S2, Canada.*
[7]*Shenzhen Institute for Quantum Science and Engineering, Southern University of Science and Technology, Shenzhen 518055, China.*
[*]*These authors contribute equally to this work.*

*Correspondence and requests for materials should be addressed to HS (email: hzsong1296@163.com), ZW (e-mail: zhmwang@gmail.com), or to QZ (e-mail: zhouqiang@uestc.edu.cn).*

## Abstract

**Heralded single photon source (HSPS) is an important way in generating genuine single photon, having advantages of experimental simplicity and versatility. However, HSPS intrinsically suffers from the trade-off between the heralded single photon rate and the single photon purity. To overcome this, one can apply multiplexing technology in different degrees of freedom to enhance the performance of HSPS. Here, by employing spectral multiplexing and active feed-forward spectral manipulating, we demonstrate a HSPS at 1.5 $\mu$m telecom-band. Our experimental results show that the spectral multiplexing effectively erases the frequency correlation of pair source and significantly improves the heralded single photon rate while keeping the $g^{(2)}(0)$ as low as 0.0006±0.0001. The Hong-Ou-Mandel interference between the heralded single photons and photons from an independent weak coherent source indicates a high indistinguishability. Our results pave a way for scalable HSPS by spectral multiplexing towards deterministic single photon emission.**

**Introduction**

Single photon source is an essential element in photonic implementations of quantum technologies, which ideally emits single photons in a pure, deterministic and indistinguishable manner that can ensure several potential applications such as security communication[1,2], exponentially enhanced calculation[3] and high accuracy measurement[4]. To make single photon emission more applicable for wide-spread deployment in quantum techniques, two different avenues have been paved, which are based on single emitters[5], and spontaneous nonlinear parametric processes, respectively. The first one is capable of emitting single photons deterministically, but needs complex fabrication processes as well as lacking of wavelength tunability. The second pathway is heralded single photon source (HSPS)[6] which is based on the photon-pairs created from spontaneous parametric down-conversion (SPDC) or spontaneous four wave mixing (SFWM). The detection of one of the photons in the photon-pairs, the heralding photon, indicates the existence of its twins, the heralded photon. Such HSPS are experimentally convenient, emission wavelength flexible, and most importantly, highly indistinguishable. However, the spontaneous nonlinear parametric processes bring inevitable probabilistic property[7]. Especially, to achieve higher single photon purity, the HSPS system has to operate at low pumping power level, which limits the heralded single photon (HSP) rate and hinders the practical application of HSPS.

To break this trade-off between the HSP rate and the single photon purity and to develop HSPS towards a higher efficiency and deterministic manner, one promising way is to multiplex a set of HSPSs into one common output[6,8,9]. In such multiplexing system, the detection of a heralding photon indicates the presence of a heralded photon at this certain mode. Many multiplexing schemes have been demonstrated, with modes in different degrees of freedom including space[10-15], time[16-18], frequency[19,20] and orbital angular momentum (OAM)[21]. Among these schemes, multiplexing HSPSs in frequency domain attracts great interest for their promising scalability as well as fixed external losses and requirement of resources[19,20]. Also, such sources are suitable for frequency encoded quantum information which are recently gained many attentions in telecommunication tasks of complex quantum state[22] and quantum storage[23,24]. In these tasks, electro-optic phase modulator (EOM) serves as a fundamental component for quantum state manipulation in frequency domain[25]. Moreover, for quantum information applications in metropolitan scale[26,27], single photons at 1.5 $\mu$m are preferred since they are within the third transmission window of optical fiber and many resources such as dense wavelength division multiplexers (DWDMs) and potential quantum memory can be utilized[28].

Although several HSPSs have been developed in this wavelength range, the one with spectral multiplexing is not yet demonstrated.

In this paper, we propose and experimentally demonstrate a HSPS based on spectral multiplexing and feed-forward control at 1.5 $\mu$m telecom-band. In our proof-of-principle experimental demonstration, broadband correlated photon-pairs are generated by the cascaded second-order harmonics generation (SHG) and SPDC processes in an integrated PPLN waveguide module[29]. Three spectral modes within the idler spectrum are defined as the heralding spectral modes while different active feed-forward spectral manipulations are operated on the signal photon side when a corresponding heralding photon is detected in spectral mode resolving way. Our experimental results show a breakthrough of trade-off between the HSP rate and the single photon purity. With continuous wave (CW) laser pumping and three modes multiplexed, the HSP rate is enhanced by near three-fold towards 3.1 kHz at low pumping power level while keeping an ultra-low $g^{(2)}(0)$ value of 0.0006±0.0001. The experimental results show that heralded single photons have a $g^{(2)}(0)$ as low as 0.0140±0.0009 with a HSP rate of 21.2 kHz. The measurement of joint spectrum intensity (JSI) reveals that the spectral manipulation at single photon level brings photons from three spectral modes to the common spectral mode. For the first time, we investigate the interference effect between spectrally multiplexed HSPS and an independent weak coherent source, achieving a Hong-Ou-Mandel (HOM) effect from independent sources with a visibility of 60.99±4.80%. This non-classical nature shows good indistinguishability and potential application in Bell-state measurement (BSM) based quantum teleportation[26,27] and linear optical quantum computing[30].

## Results

**Experimental setup**

Fig. 1 shows the experimental setup for our proposed spectrally multiplexed HSPS scheme. In this scheme, broadband photon-pairs are generated by cascaded SHG and SPDC processes in a periodically poled lithium niobite (PPLN) waveguide. Under this configuration of nonlinear processes, the wavelength of pumping photons, signal photons and heralding photons are all at 1.5 $\mu$m telecom-band, which can be compatible with optical fiber telecommunication system conveniently. After exiting from the PPLN waveguide, correlated signal photons and heralding photons with bandwidth of ∼100 GHz are selected and separated by the DWDM devices. Heralding photons with three spectral modes, labeled by their central frequencies $f_{i1}$, $f_{i2}$ and $f_{i3}$, respectively, are further filtered out by using a fiber-based narrowband DWDM with channel spacing of 12.5 GHz and the transmission bandwidth for each channel being 6.5

GHz. Here, $f_{i1}$=193.4992 THz (1549.36 nm), $f_{i2}$=193.5117 THz (1549.26 nm) and $f_{i3}$=193.5242 THz (1549.16 nm). These three modes are defined as the heralding modes. Heralding photons from each mode are individually detected by superconducting nanowire single photon detectors (SNSPDs). Detection events of those $f_{i1}$ or $f_{i3}$ photons are sent to the feed-forward logic where each detection event triggers the logic circuit, driving electronic devices to fire a corresponding frequency shifting signal towards the EOM. For signal photons, before arriving at the EOM, they are delayed by a fiber loop so as to match the timing sequence with the electronic frequency shifting signal generated by their twins. By doing so, signal photons are shifted and multiplexed into a common spectral mode (See Materials and Methods for details).

We verify the spectral property of the heralded signal photons with and without applying the frequency shifting signal, respectively. Here, signal photons are selected in frequency domain by varying the filtering window of a tunable narrowband filter (TNF) which has a bandwidth of 12.5 GHz. The coincident events between signal photons and heralding photons are counted by the coincidence logic circuit. The results are shown in Fig. 2. In Fig. 2A, without frequency shifting, it is obvious that the spectra of signal photons are separated into three spectral modes, $f_{s+}$, $f_{s0}$ and $f_{s-}$, corresponding to those heralding photons with the same mode spacing following energy conservation. Here, $f_{s+}$=195.7006 THz (1531.93 nm), $f_{s0}$=195.6881 THz (1532.03 nm) and $f_{s-}$=195.6756 THz (1532.13 nm). As frequency shifting signal applied, photons in those spectral modes $f_{s+}$ and $f_{s-}$ get frequency shifted and indistinguishably shifted into the $f_{s0}$ mode, as shown in Fig. 2B. We realize the symmetric frequency up and down shifting with an efficiency higher than 90% (See Materials and Methods for details).

**Brightness and coincidences-to-accidentals ratio (CAR)**

We characterize the source brightness, i.e. HSP rate, with experimental setup running in both multiplexing enabled and disabled cases, when CW pumping light with different powers are applied, as shown in Fig. 3A. The multiplexed HSP rate is measured by fixing the TNF at center frequency of $f_{s0}$ with multiplexing enabled, while the HSP rate of three individual spectral modes are measured by fixing the center frequency of the TNF at $f_{s+}$, $f_{s0}$ and $f_{s-}$, respectively, with multiplexing disabled.

In Fig. 3A, it can be seen that the multiplexed source has an HSP rate enhanced by a factor of 2.80±0.12 comparing with non-multiplexed individual modes in low pumping power region, i.e. less than 4 mW in our experiment. The enhancement, drawing down as pumping power increases, is intrinsically hindered because of the imperfect frequency response of electronic devices. Due to this non-ideal response, some frequency shifting signals are lost in high pumping power region (see the Supplementary Materials). Under the pump power of 16.98 mW, the HSP rate of multiplexed source reaches 23.6 kHz.

Coincidences-to-accidentals ratio (CAR) is a powerful tool to estimate how many undesired noise photons are generated together with the genuine correlated photon-pairs. It can be defined as the ratio of the coincidence count rate between signal photons and heralding events to the accidental count rate. We measure the CAR versus HSP rate for multiplexed source and three individual modes. As shown in Fig. 3B, the CAR of multiplexed source is higher than those of individual main source of noise is multi-pair events[29]. In our demonstration, a high-performance multiplexing HSPS is obtained with the CAR higher than 2000 at the HSP rate of 4 kHz. This value of multiplexed HSP rate also shows an enhancement of 2.80±0.12 when compared with individual HSP rates without multiplexing at the same CAR value. Though such enhancement would be deteriorated by the imperfect frequency response of electronic devices at high HSP rate, a CAR that higher than 100 can still be obtained.

**Purity of single photon**

We obtain the single photon purity of our spectrally multiplexed HSPS, by measuring the second-order auto-correlation function $g^{(2)}(\tau)$ through the Hanbury-Brown and Twiss (HBT) setup[31]. For the case of HSPS, the second-order auto-correlation function can be express as:

$$g^{(2)}(\tau) = \frac{C_{ABH}(\tau)H}{C_{AH}(\tau)C_{BH}(0)}, \qquad (1)$$

where $C_{ABH}$ is the coincident count of detection events from detectors $A$ and $B$ that being connected with two output ports of the beam splitter conditioned by heralding signal (i.e. three-fold coincidence); $C_{AH}$ ($C_{BH}$) represents the coincidences between the detection events from detector $A$ ($B$) and heralding signal; while $H$ is the individual count rate of heralding events. Here, $\tau$ is the electronic delay between the detection events from detector $A$ and detector $B$. A full description of $g^{(2)}(\tau)$ is shown in the inset of Fig. 3C. The $g^{(2)}(\tau)$ at zero delay, i.e. $g^{(2)}(0)$, can effectively characterize the single photon purity, and what Fig. 3C shows are the measured $g^{(2)}(0)$ under different HSP rate when multiplexing is enabled and disabled. With three spectral modes multiplexed, we obtain $g^{(2)}(0)$ as low as 0.0140±0.0009 at the measured maximum HSP rate. Comparing the $g^{(2)}(0)$ results with multiplexing on and off in Fig. 3C, we can conclude that when the system is operating at low power region, the $g^{(2)}(0)$ for the non-multiplexed HSPS can be reduced by nearly 3 times after the multiplexing is introduced. We obtain a $g^{(2)}(0)=0.0006\pm0.0001$ when the multiplexed HSP rate is 3.1 kHz, comparing with the non-multiplexed case at the same HSP rate with $g^{(2)}(0)$ near 0.0020. In the region of high HSP rate, such improvement gets weakened because of the limited response speed in the electronic system for frequency shifting signal generation. If the electronic signal generation system is designed elaborately, the improvement in single photon purity by multiplexing can be kept linear, even when the HSP rate is sufficiently high, as shown in the theoretically calculated curve in Fig. 3C.

**Joint spectral intensity and Hong-Ou-Mandel interference**

We study the spectral property of our proposed multiplexed HSPS by analyzing the joint spectral intensity (JSI) of photon-pairs and HOM interference between the multiplexed HSPS and a weak coherent single photon source. The JSI is given by the mode square of joint spectral amplitude[32] which can be measured by signal-idler photon coincidences versus $f_s$ and $f_i$, i.e. the signal frequency and the idler frequency.

Pumping the PPLN waveguide in Fig. 1 with pulsed laser, we measure the JSI (See Materials and Methods for details). Fig. 4A-C show the JSIs of photon-pairs directly output from PPLN without filtering, photon-pairs with narrowband DWDM inserted into the idler arm when multiplexing is disabled, and photon-pairs with multiplexing enabled. As shown in Fig. 4A, for the broadband nature of the PPLN waveguide, the JSI exhibits a diagonal band with strong frequency correlation. The cross-sectional width $\delta f_p$=6.4 GHz corresponds to the bandwidth of the pulsed pump. In Fig. 4B, the JSI is broken into three circle-like islands, corresponding to the three heralding modes defined by the three transmission bands of the narrowband DWDM. The circle-like shape of each island indicates that the frequency correlation is significantly canceled locally. The island value of spectral mode $f_{s+}$ is lower than other two islands, since the loss of the corresponding idler channel is 1 dB higher than the other two channels. As shown in Fig. 4C, with the spectral multiplexing turned on, the island with spectral mode $f_{s+}$ ($f_{s-}$) moves downwards (upwards) and all of the three islands enter into a common horizontal band. Within this band, the information of idler frequency almost cannot give any information about the signal frequency. This means that the multiplexed photon-pairs are spectrally uncorrelated.

To further characterize the nonclassical nature and indistinguishability of our spectrally multiplexed HSPS, we demonstrate an experiment of HOM interference between the multiplexed HSPS and an independent weak coherent single photon source. For general pair sources, the individual signal mode or heralding mode do not deliver a non-classical photon-number statistics unless extract the single photon nature by the heralding procedure. Visibility higher than 50% is used as a criterion to determine whether the non-classicality takes place in HOM interference between photons from independent sources[33]. The weak coherent source not only serves as the quasi single photon source for the HOM effect from independent sources, but also, this kind of interference between single photons and photons from weak coherence source can be used in some areas such as homodyne detection[34] and quantum circuits[35]. Here, the HOM interference experiment is carried out by launching balanced field intensity into two input ports, meaning that the mean photon number are equal when two input fields are respectively launched into two input ports of the beam splitter. As shown in Fig. 5, the visibility of two-fold coincidences without heralding procedure is 39.85%±1.63% while the visibility after heralding procedure (i.e. three-fold coincidences) reaches 60.99%±4.80% without subtracting the contribution of accidental events, indicating a non-classical nature. The three-fold coincidences are post-selected by the heralded signal while the spectral purity imperfection is taken into account. According to our theoretical calculation of the HOM interference between independent sources (see the Supplementary Materials), the theoretical upper bound of the visibility should be 64.67% when the spectrally pure single photons interfere with photons from a weak coherent source at the same mean photon number[36,37]. The genuine three-fold visibility of this HOM interference between the multiplexed HSPS and weak coherent source approximates 64.67%, indicating highly indistinguishable photons emitted from the multiplexed single photon source. If replacing the weak coherence source by a genuine single photon source, based on our theoretical model, the visibility of HOM interference is expected to be 94.3%, which is close to unity. The small degrade of visibility from unity are mainly contributed to the residual multiphoton generation events in our spectrally multiplexed HSPS, as described by the non-zero $g^{(2)}(0)$.

**Discussion**

In this paper, we demonstrate a scalable spectrally multiplexed HSPS scheme in telecom-band, yielding a significantly enhanced performance using off-the-shelf fiber-based components. The enhancement is possible to behave as our theoretical model and perfectly close to the number of multiplexing modes when the faster electronic devices utilized. Furthermore, the first implementation of HOM interference from independent sources between our multiplexed source and a weak coherent single photon source shows the evidences of non-classical nature.

In our demonstration, we used a tunable filter as the output filter for HSPS, which introduces relatively high loss. This loss, as well as losses from other fiber-based components will degrade the photons collection efficiency and heralding efficiency (see details in the Supplementary Materials). Integrating basic components including filters, EOM, beam splitters and detectors integrated on a single chip[38,39] is a good way to address the problem of component and connection losses in improving the performance of HSPS significantly.

Fig. 4C shows that spectral multiplexing could eliminate the frequency correlation in the photon-pairs created by SPDC process, and this is the basis for the single photon brightness enhancement without additional multiphoton emissions introduced in spectrally multiplexed HSPS. Other ways to generate frequency uncorrelated pure state are group velocity dispersion engineering[40,41] and nonlinear interferometer[42], in which higher collection efficiency could be realized since the loss of narrowband filters can be avoided. However, these schemes are relative difficult in designing those specific wavelengths for different spectral modes.

Furthermore, scalable multiplexing is promising in our scheme if a larger magnitude of frequency shifting is applied. The recently developed integrated lithium niobite EOM has low π-voltage about $V_\pi$=1.4V and ultra-low loss of ~0.5 dB, which is possible to reach a larger frequency shifting as well as better brightness and efficiency in spectrally multiplexed HSPS[43]. Another possible way to enlarge the magnitude of frequency shifting is to use an arbitrary waveform generator (AWG) with faster sample rate. For example, current available AWG with 100 GS/s sample rate shows a four-fold improvement comparing with our scheme. In this case, using the same way to prepare frequency shifting signal as we implemented, the magnitude of frequency shifting would be $\Delta f = 4\kappa_\perp/2V_\pi \approx 100$ GHz. Hence, considering a spectral mode spacing of 12.5 GHz in our implementation, up to 17 spectral modes can be multiplexed while consuming a bandwidth resource of only 200 GHz, which shows a good scalability.

We note that related result of spectrally multiplexed HSPS in 1.5 $\mu$m has been implemented by T. Hiemstra et al.[44]. In their scheme, the EOM is also employed as frequency converter, the spectral mode resolving detection of heralding photons is realized by a time-of-flight spectrometer while applying the linear range of sinusoidal signal for frequency shifting. In comparison, our scheme utilizes voltage ramp as frequency shifting signal, which suffer less impact from the jitter of SNSPDs and timing electronics, providing a more precise active feed-forward control. We summarize recent demonstrated works of multiplexing HSPS as well as their performance, as Table 1 shows. We highlight that our spectrally multiplexed HSPS has an ultra-high single photon purity with $g^{(2)}(0)$=0.0006±0.0001 at an HSP rate of 3.1 kHz, which means a near ideal purity that ever reported. Our spectrally multiplexed HSPS also shows high indistinguishability especially in frequency domain with a HOM visibility higher than 90%, which can benefit a lot to quantum applications, e.g. linear optical quantum computing[30]. In recent year, high performance single photon source has been intensely investigating and becoming an important building block for quantum network and quantum computing. We believe our demonstration of spectrally multiplexed HSPS is useful for developing high quality single photon source, and also useful for building quantum networks based on multiplexing schemes.

## Materials and Methods

**Detail of experimental setup**

As describe in the main text, the spectrally multiplexed HSPS is experimentally realized as following: Spectral mode resolving detection of heralding photons tells the feed-forward logic about the frequency information while feed-forward logic then gives the frequency converter, i.e. electro-optic modulator, a corresponding frequency shifting signal. Finally signal photons are shifted into the central spectral mode. Here, pumping field centered at $\lambda_p$=1540.16 nm is sent into a PPLN waveguide, which generates broadband photon-pairs with Type-0 phase matching condition. Note that both SHG and SPDC take place in the PPLN waveguide and both input and output ports of the PPLN waveguide are coupled with single mode fibers. The generated photon-pairs are selected into signal and heralding arm separately by a commercial DWDM with ~100 GHz bandwidth centered at $\lambda_s$=1531.90 nm and $\lambda_i$=1549.32 nm, respectively. The idler arm is further filtered by a fiber-based narrowband DWDM (MICS, Kylia) with channel spacing of 12.5 GHz and the transmission bandwidth of each channel being 6.5 GHz. For heralding and multiplexing procedure, those electronic signals of detection events from three different modes, $f_{i1}$, $f_{i2}$ and $f_{i3}$, are combined by the coincident logic as the heralding signal while detection events of the $f_{i1}$ or $f_{i3}$ photons generate corresponding trigger signals to the feed-forward logic. Both coincident logic and feed-forward logic are realized by a programmable time-to-digit converter (TDC, ID900, ID Quantique Corp.).

**Frequency shifting signal**

The frequency shifting signal is prepared by a series of electronic devices which are not shown in Fig. 1. For details of electronic setup in preparing the frequency shifting signal (see the Supplementary Materials). The AWG (70002A, Tektronix Corp.) operates at trigger mode, triggered by signal comes from the feed-forward logic, producing a pre-programmed pulse signal with ultra-fast pulse edge in a sampling rate of 25 GS/s. Both high-speed amplifier (S126 A, SHF Communication Technologies AG) with 25 GHz bandwidth and home-made high-voltage radio frequency (RF) transistor are used to amplify the pulse signal. The AWG signal is first amplified by the high-speed amplifier, then triggers the RF transistor to fire a reversed pulse with linear high-voltage ramp which the rising edge $\kappa_+$ and the falling edge $\kappa_-$ are symmetric. Here, $\kappa_\pm$ represents the voltage ramp of the pulse edge (see the Supplementary Materials). The pulse is further sent to the EOM (CETC Corp.), driving the EOM to modulate the phase of the single photon wave-packet as follow[19]:

$$\varepsilon(x,t) = |\varepsilon(x,t)| \exp\left[i2\pi\left(f_0 + \frac{\kappa_\pm}{2V_\pi}\right)t - ikx\right], \quad (2)$$

where the single photon wave-packet $\varepsilon(x,t)$ is depict by the slowly-varying-envelope approximation while $V_\pi$ is the π-voltage of the EOM and $\kappa_\pm$ is the slope of the voltage signal. The linearly varying time-dependent change to the phase of the photon yields a frequency shifting with the shifting magnitude of $\Delta f = \kappa_\pm/2V_\pi$. Here, the ultra-fast pulse edge $\kappa_+$ ($\kappa_-$) leads to frequency up (down) conversion. To realize precise frequency shifting, the trigger signals generated by detection events of $f_{i1}$ and $f_{i3}$ photons are combined by the feed-forward logic, where a certain electronic delay is added to the $f_{i1}$ signals to compensate the time difference between pulse edges $\kappa_+$ and $\kappa_-$ in the same pulse. The trigger mode of AWG has a low pass filter like frequency response. In higher pumping power region, the frequency shifting signal has higher possibility to lost. This unwanted mechanism limits the enhancement of our multiplexed HSPS. But it should be noted that it is available to correct this limit using faster commercial devices. As Fig. 3 shows, the ideal performance will follow the dotted line of theoretical calculation. Detailed characterizations of this frequency response is shown in the Supplementary Materials.

**Measurement of joint spectral intensity**

In the scheme of JSI measurement, we add another TNF in heralding arm before the narrowband DWDM (not shown in Fig. 1). The pulse laser pump is prepared by an intensity modulator (IM, CETC Corp.) that modulates CW laser (PPCL550, PURE Photonics Corp.) with a programmed RF signal generated from AWG at sampling rate 25 GS/s, which has a ~65 ps width and 500MHz repetition rate. The peak power of this pulse laser is 25 mW. With two TNFs (12.5 GHz bandwidth) placing separately in both heralding and signal arms, the coincidences are measured by swapping TNFs' central wavelength respectively from 193.4867 THz to 193.5367 THz for heralding arm and from 195.6631 THz to 195.7131 THz for signal arm in a step of 2.5 GHz.

**Hong-Ou-Mandel interference**

We use the aforementioned pulse laser for pumping and generating multiplexed heralded single photons while the weak coherent source is synchronized with the pumping laser and has the same pulse width and repetition rate. The wavelength of the weak coherent source is centered at 1532.03 nm, matching the output wavelength of multiplexed HSPS. The signal arm of the multiplexed source and weak coherent source is coupled by a 2×2 beam splitter with both output ports being sent to SNSPDs (P-CS-6, PHPTEC Corp.). By varying the optical tunable delay line and hence changing the relative delay between the two sources, HOM effect can be observed (see the Supplementary Materials). Two-fold and three-fold coincidences are realized by measuring the coincident count between two beam splitter outputs without and with being conditioned by heralding signal, respectively.

## References

1. Gisin, N., Ribordy, G., Tittel, W. & Zbinden, H. Quantum cryptography. *Rev. Mod. Phys.* **74**, 145 (2002).
2. Gisin, N. & Thew, R. Quantum communication. *Nat. Photon* **1**, 165 (2007).
3. O'brien, J. L. Optical quantum computing. *Science* **318**, 1567–1570 (2007).
4. Giovannetti, V., Lloyd, S. & Maccone, L. Quantum metrology. *Phys. Rev. Lett.* **96**, 010401 (2006).
5. Aharonovich, I., Englund, D. & Toth, M. Solid-state single-photon emitters. *Nat. Photon* **10**, 631 (2016).
6. Meyer-Scott, E., Silberhorn, C. & Migdall, A. Single-photon sources: Approaching the ideal through multiplexing. *Rev. Sci. Instruments* **91**, 041101 (2020).
7. Christ, A. & Silberhorn, C. Limits on the deterministic creation of pure single-photon states using parametric down-conversion. *Phys. Rev. A* **85**, 023829 (2012).
8. Migdall, A. L., Branning, D. & Castelletto, S. Tailoring single-photon and multiphoton probabilities of a single-photon on-demand source. *Phys. Rev. A* **66**, 053805 (2002).
9. Pittman, T., Jacobs, B. & Franson, J. Single photons on pseudodemand from stored parametric down-conversion. *Phys. Rev. A* **66**, 042303 (2002).
10. Ma, X.-s., Zotter, S., Kofler, J., Jennewein, T. & Zeilinger, A. Experimental generation of single photons via active multiplexing. *Phys. Rev. A* **83**, 043814 (2011).
11. Collins, M. J. et al. Integrated spatial multiplexing of heralded single-photon sources. *Nat. Commun.* **4**, 1–7 (2013).
12. Xiong, C. et al. Bidirectional multiplexing of heralded single photons from a silicon chip. *Opt. Lett.* **38**, 5176–5179 (2013).
13. Francis-Jones, R. J., Hoggarth, R. A. & Mosley, P. J. All-fiber multiplexed source of high-purity single photons. *Optica* **3**, 1270–1273 (2016).
14. Mendoza, G. J. et al. Active temporal and spatial multiplexing of photons. *Optica* **3**, 127–132 (2016).

15. Meany, T. et al. Hybrid photonic circuit for multiplexed heralded single photons. *Laser & Photonics Rev.* **8**, L42–L46 (2014).

16. Kaneda, F. et al. Time-multiplexed heralded single-photon source. *Optica* **2**, 1010–1013 (2015).

17. Xiong, C. et al. Active temporal multiplexing of indistinguishable heralded single photons. *Nat. Commun.* **7**, 1–6 (2016).

18. Kaneda, F. & Kwiat, P. G. High-efficiency single-photon generation via large-scale active time multiplexing. *Sci. Advances* **5**, eaaw8586 (2019).

19. Puigibert, M. G. et al. Heralded single photons based on spectral multiplexing and feed-forward control. *Phys. Rev. Lett.* **119**, 083601 (2017).

20. Joshi, C., Farsi, A., Clemmen, S., Ramelow, S. & Gaeta, A. L. Frequency multiplexing for quasi-deterministic heralded single-photon sources. *Nat. Commun.* **9**, 1–8 (2018).

21. Liu, S.-l. et al. Multiplexing heralded single photons in orbital-angular-momentum space. *Phys. Rev. A* **100**, 013833 (2019).

22. Kues, M. et al. On-chip generation of high-dimensional entangled quantum states and their coherent control. *Nature* **546**, 622–626 (2017).

23. Sinclair, N. et al. Spectral multiplexing for scalable quantum photonics using an atomic frequency comb quantum memory and feed-forward control. *Phys. Rev. Lett.* **113**, 053603 (2014).

24. Seri, A. et al. Quantum storage of frequency-multiplexed heralded single photons. *Phys. Rev. Lett.* **123**, 080502 (2019).

25. Wright, L. J., Karpinski, M., Söller, C. & Smith, B. J. Spectral shearing of quantum light pulses by electro-optic phase modulation. *Phys. Rev. Lett.* **118**, 023601 (2017).

26. Valivarthi, R. et al. Quantum teleportation across a metropolitan fibre network. *Nat. Photon* **10**, 676 (2016).

27. Sun, Q.-C. et al. Quantum teleportation with independent sources and prior entanglement distribution over a network. *Nat. Photon* **10**, 671–675 (2016).

28. Saglamyurek, E. et al. A multiplexed light-matter interface for fibre-based quantum networks. *Nat. Commun.* **7**:11202 (2016).

29. Zhang, Z. et al. High-performance quantum entanglement generation via cascaded second-order nonlinear processes. Preprint at https://arXiv.org/abs/2102.07146 (2021).

30. Knill, E., Laflamme, R. & Milburn, G. J. A scheme for efficient quantum computation with linear optics. *Nature* **409**, 46–52 (2001).

31. Brown, R. H. & Twiss, R. Q. Correlation between photons in two coherent beams of light. *Nature* **177**, 27–29 (1956).

32. Kumar, R., Ong, J. R., Savanier, M. & Mookherjea, S. Controlling the spectrum of photons generated on a siliconnanophotonic chip. *Nat. Commun.* **5**, 1–7 (2014).

33. Li, X., Yang, L., Cui, L., Ou, Z. Y. & Yu, D. Observation of quantum interference between a single-photon state and athermal state generated in optical fibers. *Opt. Exp.* **16**, 12505–12510 (2008).

34. Lvovsky, A. I. & Raymer, M. G. Continuous-variable optical quantum-state tomography. *Rev. Mod. Phys.* **81**, 299 (2009).

35. Pittman, T. B., Jacobs, B. C. & Franson, J. D. Experimental demonstration of a quantum circuit using linear optics gates. *Phys. Rev. A* **71**, 032307 (2005).

36. Fearn, H. & Loudon, R. Theory of two-photon interference. *J. Opt. Soc. Am. B* **6**, 917–927 (1989).

37. Rarity, J., Tapster, P. & Loudon, R. Non-classical interference between independent sources. *J. Opt. B: Quantum Semiclassical Opt.* **7**, S171 (2005).

38. Hu, Y. et al. Reconfigurable electro-optic frequency shifter. Preprint at https://arXiv.org/abs/2005.09621 (2020).

39. Marpaung, D. et al. Integrated microwave photonics. *Laser & Photonics Rev.* **7**, 506–538 (2013).

40. Grice, W. P., U'Ren, A. B. & Walmsley, I. A. Eliminating frequency and space-time correlations in multiphoton states. *Phys. Rev. A* **64**, 063815 (2001).

41. Cui, L., Li, X. & Zhao, N. Minimizing the frequency correlation of photon pairs in photonic crystal fibers. *N. J. Phys.* **14**, 123001 (2012).

42. Su, J. et al. Versatile and precise quantum state engineering by using nonlinear interferometers. *Opt. Exp.* **27**, 20479–20492 (2019).

43. Wang, C. et al. Integrated lithium niobate electro-optic modulators operating at cmos-compatible voltages. *Nature* **562**, 101–104 (2018).

44. Hiemstra, T. et al. Pure single photons from scalable frequency multiplexing. *Phys. Rev. Appl.* **14**, 014052 (2020).

**Acknowledgements**

This work is supported by National Key R&D Program of China (Grant Nos. 2018YFA0307400, 2018YFA0306102, 2019YFB2203400, 2017YFB0405100, 2017YFA0304000); National Natural Science Foundation of China (Grant Nos. 61775025, U19A2076, 61405030, 12074058, 91836102, 61704164, 62075034, 62005039); Sichuan Science and Technology Program (Grant Nos. 2018JY0084).

**Author contributions**

HY and CY contributed equally to this work. QZ convinced and supervised the project. HY, CY, RZ and ZZ developed the experimental setup, and performed the experiments. HY and CY analyzed the experimental results and performed the numerical studies with helps from ZW and HS. The SNSPDs were designed and fabricated by HL and LY. All authors contributed to the writing of the manuscript.

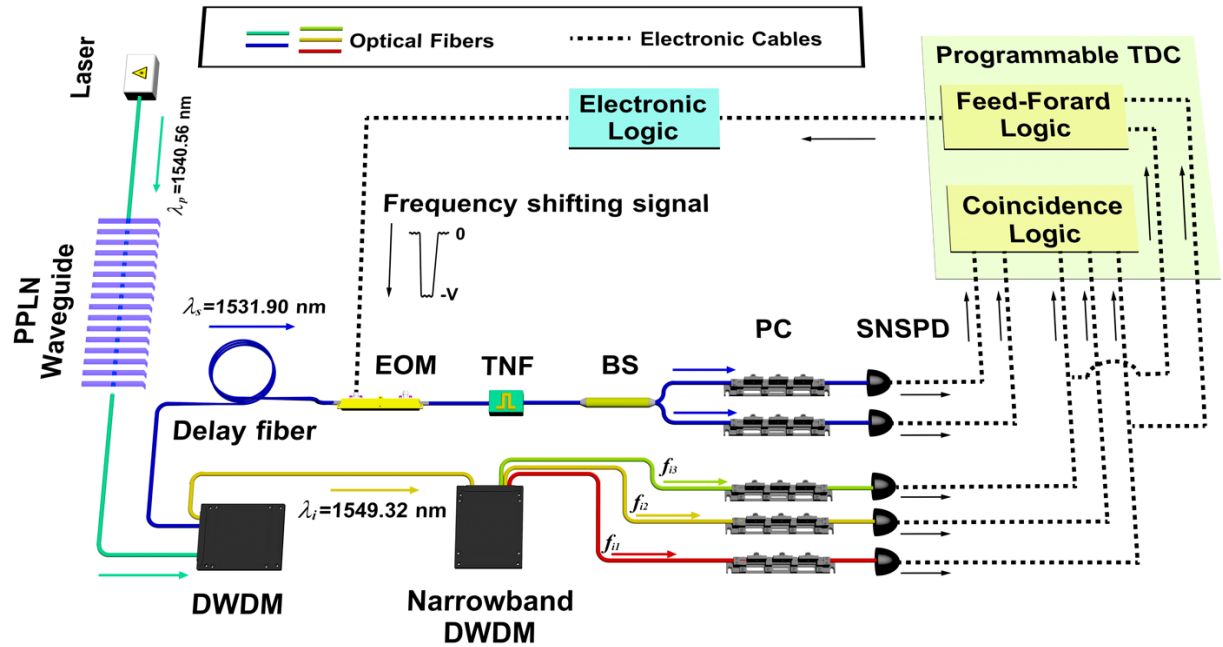

**Figure 1. Experimental setup for spectrally multiplexed HSPS.** PPLN: Periodically-Poled Lithium Niobate; DWDM: Dense Wavelength Divided Multiplexer; EOM: Electro-Optic phase Modulator; TNF: Tunable Narrowband Filter; BS: Beam Splitter; PC: Polarization Controller; TDC: Time-to-Digit Convertor; SNSPD: Superconducting Nanowire Single Photon Detector.

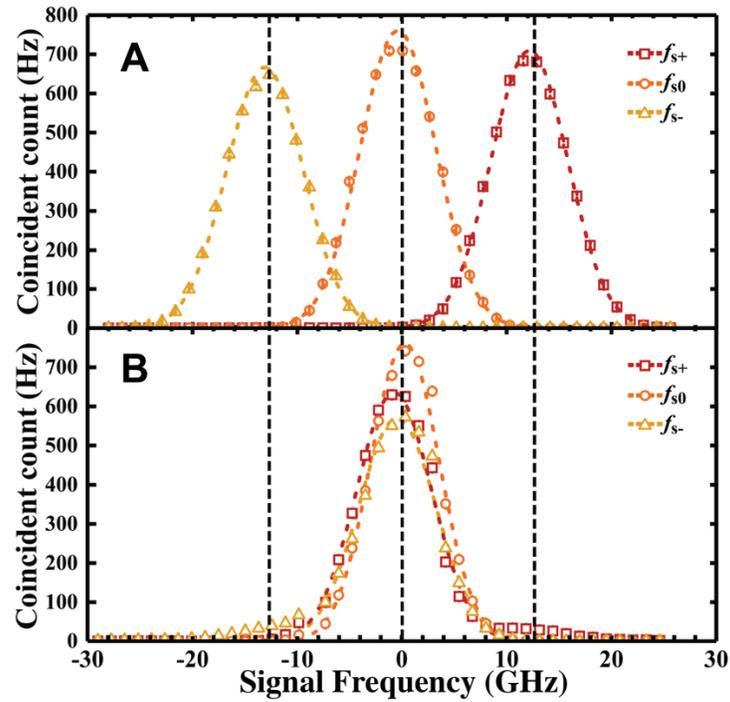

**Figure 2. Coincidence count rates without and with frequency shifting.** **(A)** and **(B)** are the measured coincidence count rates among three individual spectral modes without and with spectral multiplexing, respectively, under the CW pumping power of 4 mW. The horizontal axis corresponds to the relative frequency difference with respect to $f_{s0}$. Red rectangles, orange circles and yellow triangles represent the coincident count rate of single photons to heralding photons at spectral modes $f_{s+}$, $f_{s0}$, and $f_{s-}$, respectively. Dashed curves are fitted by Gaussian function which describe the frequency response of individual channels from the narrowband DWDM.

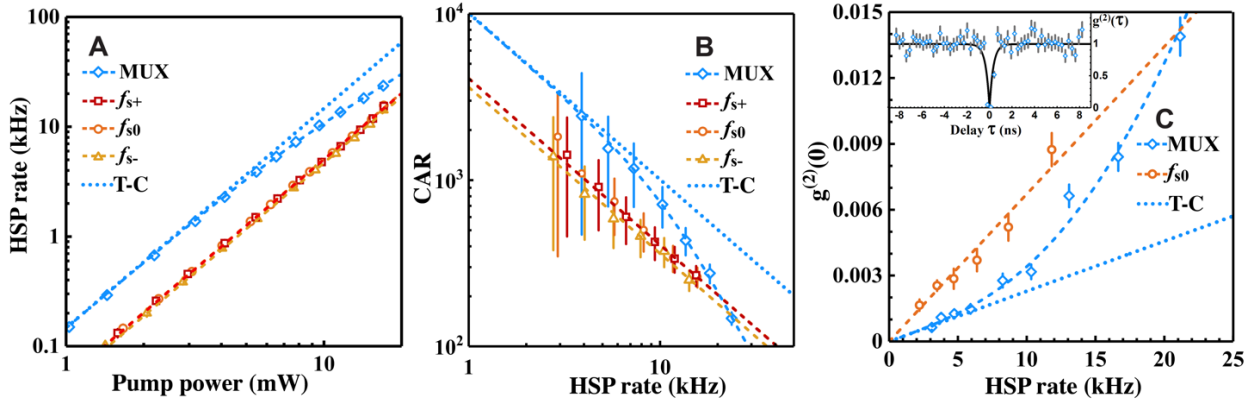

**Figure 3. Experimental results from multiplexing of three frequency modes.** Blue diamonds are multiplexed source and red rectangles, orange circles and yellow triangles represent spectral modes $f_{s+}$, $f_{s0}$, and $f_{s-}$, respectively. Dashed curves are fitted according the theory model of our experiment. Dotted line is the theoretical calculation under the condition of perfect frequency reaction of electronic devices. (**A**) HSP rate versus pump power. The HSP rate of multiplexed source is 2.80±0.12 times larger than that of individual spectral modes at low pump power. The multiplexed HSP rate reaches 23.6 kHz under the pumping power of 16.98 mW; (**B**) CAR versus HSP rate. The CAR of multiplexed source shows an improved performance than that of individual HSPS under fixed HSP rate. At low count rate, the multiplexed source has a CAR exceeding 2000 along with 2.80±0.12 times enhancement while remains high at 100 for large HSP rates. (**C**) $g^{(2)}(0)$ versus HSP rate. We select the spectral mode $f_{s0}$ as a contrast to its multiplexed counterpart. It shows that the $g^{(2)}(0)$ of individual source improves nearly 3 times rapider than the multiplexed one in low HSP rate region as well as theoretically. Inset: heralded single photon auto-correlation at HSP rate of 21.1 kHz. The black curve represents the $g^{(2)}(\tau)$ fitting. The antibunching dip is of 0.0140±0.0009. MUX: spectral multiplexing; T-C: theoretical calculation; HPS: heralded photon rate. Error bars are estimated using Poisson statistics.

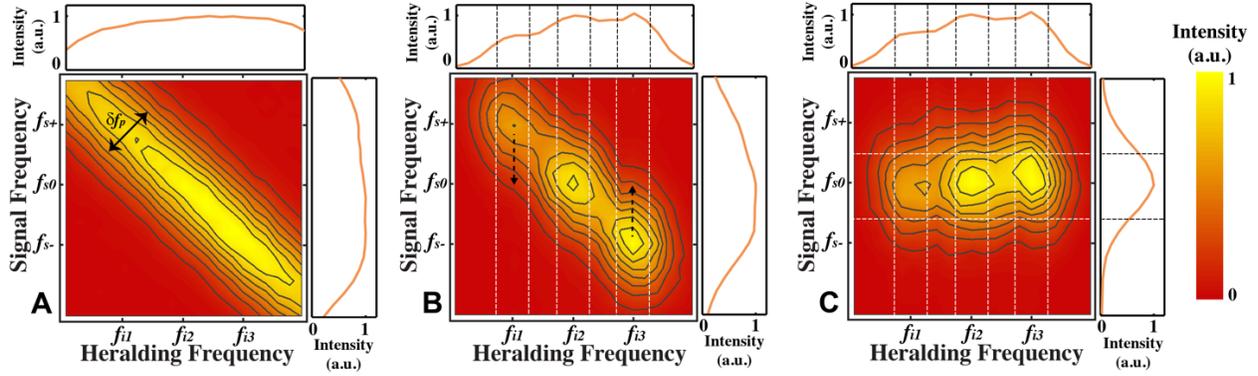

**Figure 4. Joint spectrum intensity (JSI) between heralding and signal photons in the $f_s$–$f_i$ plane.** Three JSIs shown here are measured by placing two TNFs in the signal arm and heralding arm, respectively, and counting the coincidences by sweeping frequencies of $f_s$ and $f_i$. The top and right graphs for each JSI represent the marginal intensity distribution of heralding and signal side, respectively. The vertical and horizontal white (black) dashed lines depict the bandwidth of filters for the heralding and signal arms, respectively. Here, $f_{i1}$=193.4992 THz (1549.36 nm), $f_{i2}$=193.5117 THz (1549.26 nm) and $f_{i3}$=193.5242 THz (1549.16 nm) while $f_{s+}$=195.7006 THz (1531.93 nm), $f_{s0}$=195.6881 THz (1532.03 nm) and $f_{s-}$=195.6756 THz (1532.13 nm). (**A**) Photon-pairs directly out of PPLN waveguide without filtering. The diagonal band exhibits a highly frequency correlation. (**B**) Photon-pairs with narrowband DWDM applied in heralding arm when multiplexing disabled. The JSI is broken into islands with three spectral mode corresponding to the narrowband DWDM. The circle-like island from each spectral mode exhibits a good mode purity. Note that spectral mode $f_{s+}$ has an additional 1 dB loss from idler arm. **c** Photon-pairs with multiplexing enabled. The islands of spectral modes $f_{s+}$ and $f_{s-}$ are shifted into the central spectral mode.

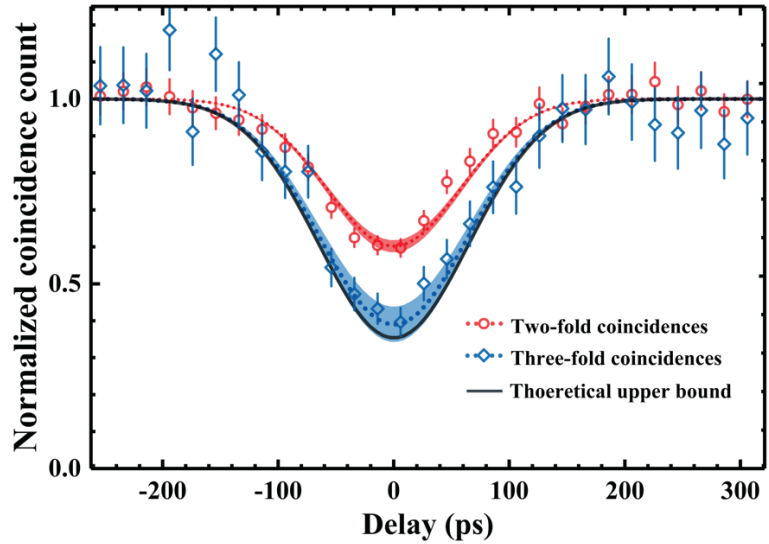

**Figure 5. HOM interference between the multiplexed source and the weak coherent source.** Red circles are two-fold coincidences measured without heralding procedure. Blue diamonds are three-fold coincidences measured under the condition by heralding procedure without subtracting the contribution of multiphoton event. Both two-fold and three-fold coincidences are fitted by gaussian function with red and blue curves and have the visibility of 39.85%±1.63% and 60.99%±4.80%, respectively. Both two-fold and three-fold visibility are 1000-time Gaussian fitting results with the Monte Carlo method. The red and blue bands represent the fitting variances of two-fold and three-fold HOM effects, respectively. The black solid curve represents the theoretically calculated upper bound of the three-fold HOM interference at 64.67%. Error bars are estimated using Poisson statistics.

**Table 1. Summary of multiplexing HSPS performance among previous demonstrations.** Sw. Num.: switch number; EF: enhancement factor; BBO: β-barium-borate crystal; PCW: photonic crystal waveguide; PLZT: lead lanthanum zirconium titanate switch; BiBO: bismuth barium borate; PCF: photonic crystal fiber; SNOW: silicon nanowire optical waveguide; BS-FWM: Bragg scattering four-wave mixing; PPKTP: periodically poled potassium titanyl phosphate; SLM: spatial light modulator.

| Author/Year | Medium | Platform | Degree | λ (nm) | Mode Num. | Switch | Sw. Num. | $g^{(2)}(0)$ | Rate (kHz) | EF |
|---|---|---|---|---|---|---|---|---|---|---|
| Ma et al.[10]/2011 | BBO/SPDC | Free-space | Space | 1550 | 4 | EOM | 3 | 0.08 | 0.7 | 3.61 |
| Collins et al.[11]/2013 | PhCW/SFWM | Integrated | Space | 1550 | 2 | PLZTS | 1 | 0.19 | 0.02 | 1.63 |
| Xiong et al.[12]/2013 | PhCW/SFWM | Integrated | Space | 1550 | 2 | PLZTS | 1 | - | 0.06 | 1.51 |
| Meany et al.[15]/2014 | PPLN/SPDC | Integrated | Space | 1550 | 4 | PLZTS | 3 | - | ~0.07 | 3 |
| Kaneda et al.[16]/2015 | BiBO/SPDC | Free-space | Time | 710 | 30 | Pockels cell | 1 | 0.479 | 19.3 | ~6 |
| F.-Jones et al.[13]/2016 | PCF/SFWM | Fiber | Space | 1550 | 2 | Optical switch | 1 | 0.05 | 0.5 | 1.75 |
| Xiong et al.[17]/2016 | SNOW/SFWM | Integrated | Time | 1545 | 4 | PLZTS | 3 | - | 0.6 | 2 |
| Mendoza et al.[14]/2016 | PPLN/SPDC | Fiber | Space & time | 1547 | 8 | Optical switch | 3 | - | 0.3 | 2.14 |
| Puigibert et al.[19]/2017 | PPLN/SPDC | Fiber | Frequency | 795 | 3 | EOM | 1 | 0.06 | 0.3 | ~1 |
| Joshi et al.[20]/2018 | PPLN/SPDC | Fiber | Frequency | 1280 | 3 | BS-FWM | 1 | 0.07 | 23 | 2.2 |
| Liu et al.[21]/2019 | PPKTP/SPDC | Free-space | OAM | 1550 | 3 | SLM | 1 | 0.048 | ~4.2 | 1.47 |
| Kaneda et al.[18]/2019 | PPKTP/SPDC | Free-space | Time | 1590 | 40 | Pockels cell | 1 | 0.007 / 0.088 / 0.269 | 25.5 / 206 / 334 | 27.9 / 18.7 / 9.7 |
| Hiemstra et al.[44]/2020 | PPKTP/SPDC | Fiber | Frequency | 1565 | ~3 | EOM | 1 | ~0.01 | ~12 | 2.7 |
| **Our work** | **PPLN/SPDC** | **Fiber** | **Frequency** | **1532** | **3** | **EOM** | **1** | **0.0006** | **23.6** | **~2.8** |

# Supplementary Materials for Spectrally multiplexed heralded single photon source at telecom-band

Hao Yu[1,6,*], Chenzhi Yuan[1,*], Ruiming Zhang[1], Zichang Zhang[1], Hao Li[5], You Wang[1,3], Guangwei Deng[1], Lixing You[5], Haizhi Song[1,3], Zhiming Wang[1,7], Guang-Can Guo[1,4], and Qiang Zhou[1,2,4]

[1]*Institute of Fundamental and Frontier Sciences, University of Electronic Science and Technology of China, Chengdu 610054, China.*
[2]*Yangtze Delta Region Institute (Huzhou) & School of Optoelectronic Science and Engineering, University of Electronic Science and Technology of China, Huzhou 313001, China.*
[3]*Southwest Institute of Technical Physics, Chengdu 610041, China.*
[4]*CAS Key Laboratory of Quantum Information, University of Science and Technology of China, Hefei 230026, China.*
[5]*Shanghai Institute of Microsystem and information Technology, Chinese Academy of Sciences, Shanghai 200050, China.*
[6]*Institut national de la recherche scientifique - Centre Énergie, Matériaux et Télécommunications (INRS-EMT), Varennes J3X 1S2, Canada.*
[7]*Shenzhen Institute for Quantum Science and Engineering, Southern University of Science and Technology, Shenzhen 518055, China.*
[*]*These authors contribute equally to this work.*

*Correspondence and requests for materials should be addressed to HS (email: hzsong1296@163.com), ZW (e-mail: zhmwang@gmail.com), or to QZ (e-mail: zhouqiang@uestc.edu.cn).*

**Supplementary Text I**

**Electronic devices for generating frequency shifting signal**

In our scheme of spectrally multiplexed heralded single photon source (HSPS), the frequency shifting signal is generated from a series of electronic devices, see Fig. S1(A). The key component in the circuit is a high-voltage and large-bandwidth radio frequency (RF) transistor that acts as an inverting electrical amplifier. The detection event from the $f_{i1}$ or $f_{i3}$ photon can generate an electrical pulse by a feed-forward logic circuit, which triggers the generation of a high-voltage pulse with very fast falling and raising edges, see Fig. S1(B). The enhancement versus heralding rate of the multiplexed source system is shown in Fig. S1(C). The deterioration of enhancement is mainly contributed to the poor frequency response of the arbitrary waveform generator (AWG). Here, the AWG runs in trigger mode which each input signal that comes from the feed-forward logic generates a corresponding pulse in the AWG. The trigger mode of AWG has a low-pass like frequency response, which we model as a low order Butterworth filter with a 3 dB bandwidth of 1.2 MHz. The feed-forward signal and therefore the frequency shifting signal has a higher possibility to loss under higher heralding rate. In this way, $f_{s+}$ and $f_{s-}$ photons have less possibility to be shifted into the center spectral mode and ultimately degrade the enhancement of multiplexed source. Similarly, the heralding efficiency are also affected by this mechanism as the inset of Fig. S1(C) shows. The data points are in agreement with the theoretical prediction. Here, we note that the degradation of enhancement is only affected by the inner mechanism of AWG. Using currently off-the-shelf electronic devices, the performance of our multiplexed source will go as the theoretical calculation curve (see main text Fig. 3).

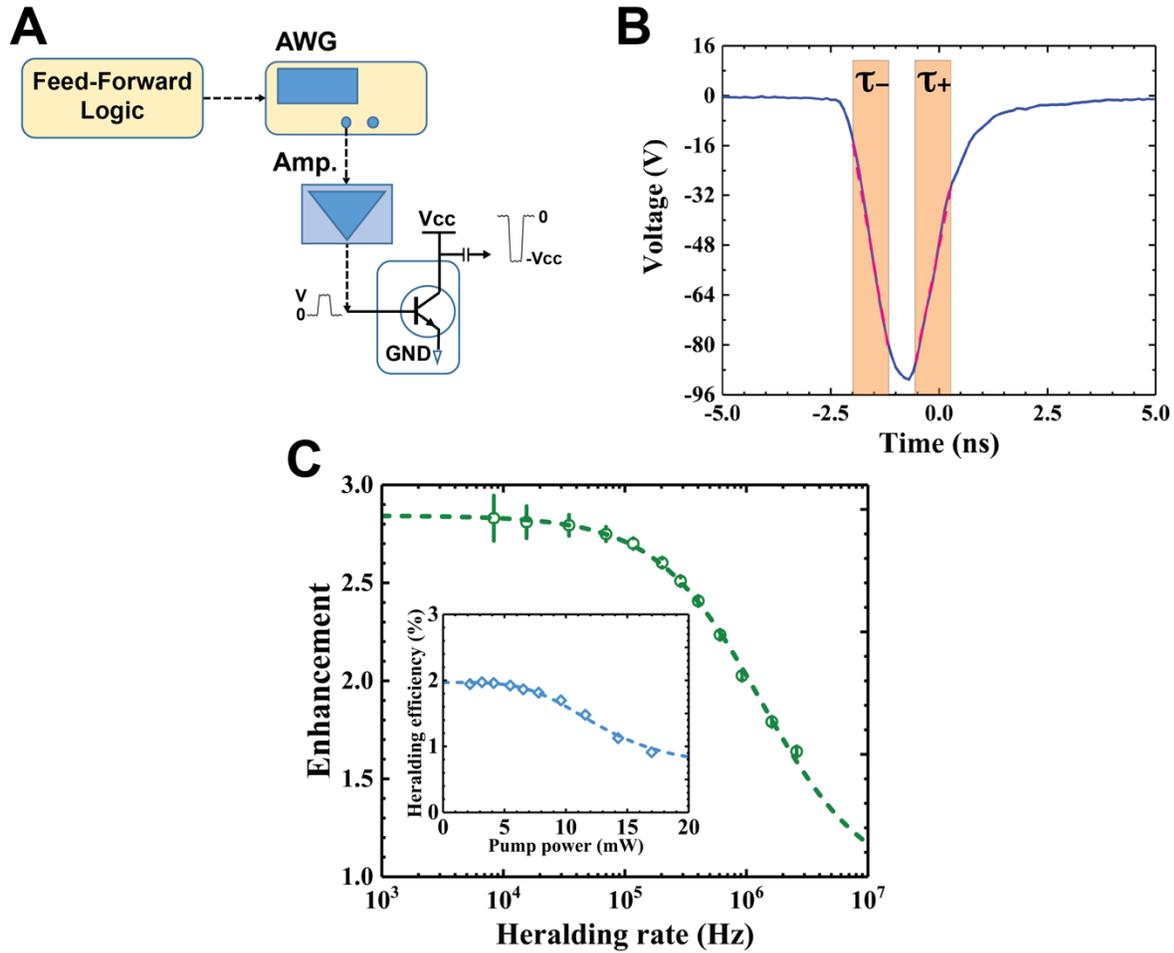

**Figure S1.** **(A)** Experimental setup for generation of frequency shifting signals. AWG: Arbitrary Waveform Generator; Amp.: Microwave Amplifier; **(B)** Typical electronic signal for frequency shifting; **(C)** Enhancement of the multiplexed source versus heralding rate. Inset: Heralding efficiency of the multiplexed source versus pump power. Green circles and blue diamonds are measured data while green and blue dashed curves are the result of theoretical calculation. Error bars are estimated using Poisson statistics.

**Supplementary Text II**

**Measurement of Hong-Ou-Mandel interference**

The setup of our Hong-Ou-Mandel (HOM) interference from individual sources are shown in Fig. S2. Here, multiplexing heralded single photon source and weak coherent source are used to perform the HOM interference. The pulsed laser is synchronized with the photon-pair source with 6.4 GHz pulse width and 500 MHz repetition rate. The HOM effect only occurs when two input photons are indistinguishable. Thus, the wavelength of the two sources are both centered at 1532.03 nm ($f_{s0}$) while input photons are both aligned to the same polarization with the help of polarization controllers. Noting that multiplexing is actively enabled when performing HOM interference (feed-forward logic and frequency shifting signal are not shown in Fig. S2). The output ports of the beam splitter are sent to two superconducting nanowire single photon detectors (SNSPDs) while the detection signals are collected and analyzed by the coincident logic. Beside for feed-forward control, the heralding signals are also combined and sent to the coincident logic. The heralding signal is a conditional trigger for three-fold coincidences. The coincident logic can simultaneously analyze the coincident events of the two beam splitter output ports with and without the trigger of heralding signal, and hence we can both measure two-fold and three-fold coincidences at the same time. Therefore, by varying the optical tunable delay line and hence changing the relative delay between the two sources, HOM effect can be observed.

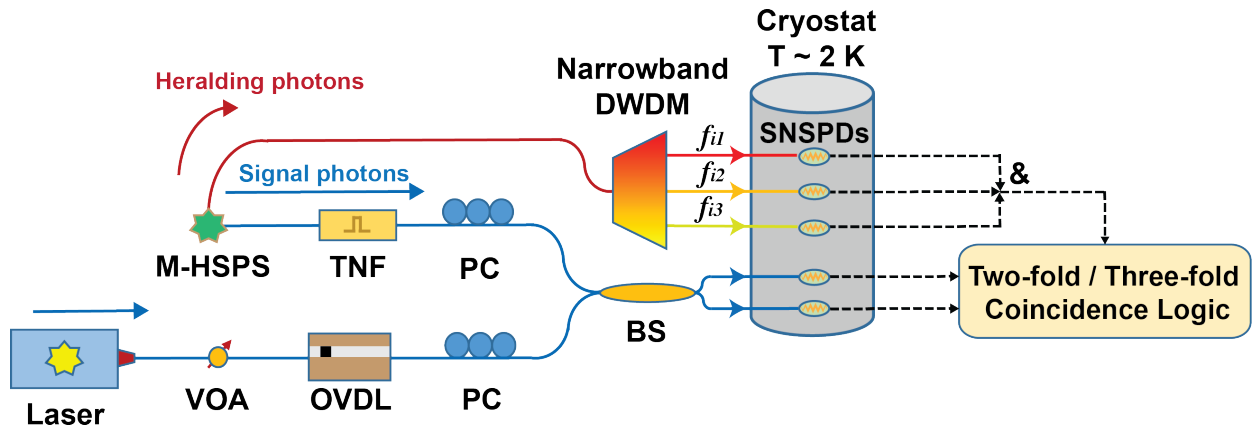

**Figure S2.** Experimental setup of HOM interference from individual sources (the feed-forward logic and frequency converter are not shown here). M-HSPS: Multiplexed Heralded Single Photon Source; TNF: Tunable Narrowband Filter; VOA: Variable Optical Attenuator; OVDL: Optical Variable Delay Line; PC: Polarization Controller; BS: Beam Splitter; Narrowband DWDM: Narrowband Dense Wavelength Divided Multiplexer; SNSPD: Superconducting Nanowire Single Photon Detector.

**Supplementary Text III**

**Hong-Ou-Mandel interference visibility**

To understand the result of main text Fig. 5, we start with analyzing the ideal HOM effect from independent sources. First, let us consider the 50:50 beam splitter with two input ports and two output ports, where field operators for input ports are represented as $\hat{a}_1$, $\hat{a}_2$ while those for output ports are represented as $\hat{a}_3$, $\hat{a}_4$. We can express these four field operators as:

$$\hat{a}_3 = (\hat{a}_1 + \hat{a}_2)/\sqrt{2}, \tag{S1}$$

$$\hat{a}_4 = (\hat{a}_1 - \hat{a}_2)/\sqrt{2}, \tag{S2}$$

which we can calculate the output coincidence probability by considering their input fields. Here, the output coincidence probability $P$ can be described as[1,2]:

$$P \propto \langle \hat{a}_3^\dagger(t)\hat{a}_4^\dagger(t')\hat{a}_4(t')\hat{a}_3(t)\rangle = \frac{1}{4}\begin{pmatrix}\langle \hat{a}_1^\dagger(t)\hat{a}_1^\dagger(t')\hat{a}_1(t')\hat{a}_1(t)\rangle + \langle \hat{a}_2^\dagger(t)\hat{a}_2^\dagger(t')\hat{a}_2(t')\hat{a}_2(t)\rangle \\ +\langle \hat{a}_1^\dagger(t)\hat{a}_2^\dagger(t')\hat{a}_2(t')\hat{a}_1(t)\rangle + \langle \hat{a}_2^\dagger(t)\hat{a}_1^\dagger(t')\hat{a}_1(t')\hat{a}_2(t)\rangle \\ -\langle \hat{a}_2^\dagger(t)\hat{a}_1^\dagger(t')\hat{a}_2(t')\hat{a}_1(t)\rangle - \langle \hat{a}_1^\dagger(t)\hat{a}_2^\dagger(t')\hat{a}_1(t')\hat{a}_2(t)\rangle\end{pmatrix}, \tag{S3}$$

where in the right-hand side of this equation, the first two terms are auto-correlation function for the two input fields while the third and the fourth terms are cross-correlation between these two fields. The last two terms serve as interference terms which leads interference effect. We can further derive a simpler description of Eqs. (S3) as follow:

$$P \propto \frac{1}{4}\left(g_1^{(2)}(0)\bar{n}_1^2 + g_2^{(2)}(0)\bar{n}_2^2 + 2\bar{n}_1\bar{n}_2 - 2\bar{n}_1\bar{n}_2 I_{12}\right), \tag{S4}$$

where $\bar{n}_1$ and $\bar{n}_2$ are the mean photon number for the input fields of port1 and port2, respectively, and $I_{12} = \langle \hat{a}_1^\dagger(t)\hat{a}_2^\dagger(t')\hat{a}_1(t')\hat{a}_2(t)\rangle/\bar{n}_1\bar{n}_2$ represents the interference factor which ranges from 0 to 1 and depends on the temporal overlap of the two input fields. When two input field are completely overlapped, i.e. $t = t'$, the interference factor shows $I_{12} = 0$, which corresponding to the dip of HOM effect; When two input fields are temporally separated, the interference factor becomes 1, i.e. $I_{12} = 1$, and hence corresponds to the wing.

Now, considering the input fields $\hat{a}_1$ and $\hat{a}_2$ are ideally a thermal field and a coherent field, respectively, as the two-fold coincidence we demonstrate. We can derive the visibility of two-fold coincidence from Eq.(S3):

$$V_{two-fold} = \frac{2}{2\frac{\bar{n}_1}{\bar{n}_2} + \frac{\bar{n}_2}{\bar{n}_1} + 2}, \tag{S5}$$

In our implementation, we have a balanced input fields, which means that the mean photon number for the two input fields are equal, i.e. $\bar{n}_1 = \bar{n}_2$. That leads to a theoretical upper bound of two-fold visibility as $V_{two-fold} = 40\%$.

Next, we consider the input field $\hat{a}_1$ and $\hat{a}_2$ are ideally a single photon field and a coherent field, respectively, and hence the three-fold coincidence can also express as:

$$V_{three-fold} = \frac{2}{\frac{\bar{n}_2}{\bar{n}_1} + 2}. \tag{S6}$$

Again, in a balanced input with $\bar{n}_1 = \bar{n}_2$, it can be obtained that a theoretical three-fold visibility is upper bounded to $V_{three-fold} = 66.67\%$. Considering a 97% degrade from the mismatching of independent bandwidths, the real theoretical upper bound is 64.67%. Noting that in the configuration of coincidence logic, the single photon coincident events are post-selected by heralding signal with gate logic, which the ratio of mean photon number between both input fields remains intact.

In HOM interference, the premise for ideal non-classical interference is that the photon participating in the interference is spectrally pure. The imperfection of single photon spectral purity should be taken into consideration when revealing the non-classical nature of the three-fold coincidence. For the imperfect spectral purity, we should take into account the unwanted external broadening mechanism. In our experiment, the bandwidth of the Gaussian pumping pulses is 6.4 GHz. Considering the SHG process and heralding procedure, the heralded signal photons have a broadened external bandwidth of 11.2 GHz, which is a convolution result of pump photon bandwidth and heralding photon bandwidth. Such mismatch between the internal and external bandwidth reduces the spectral purity of the single photon and shows an 81.16% deterioration of visibility[3] as Fig. S3 shows. This degrade can be further improved by applying a narrower output filter for the HSPS. After correcting the spectral purity imperfection, the visibility of three-fold coincidences ends up with 60.99%±3.50%.

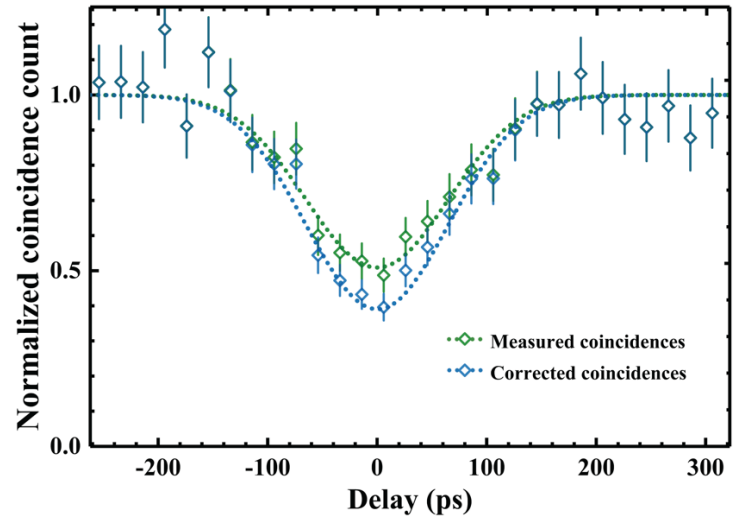

**Figure S3.** Three-fold coincidences before (green diamonds) and after (blue diamonds) correction. Error bars are estimated using Poisson statistics.

**Supplementary Text IV**

**Losses and efficiency**

The implementation of multiplexed HSPS has many fiber-based optical components, which introduce relatively large losses causing inefficient photon collection on both of heralding and heralded sides. We measure the overall component losses and efficiencies of the signal arm and heralding arm in three frequency modes, as shown in Table S1 and Table S2. The total transmission efficiencies are the direct measurement of the component losses while the collection efficiencies are calculated by Klyshko method[4]. Here, the transmission efficiency in signal arm is in accordance with its collection efficiency. The mismatch between transmission efficiency and collection efficiency in heralding arm is because of the difference of filtering bandwidths between signal and idler arms, which reducing the collection efficiency by a ratio of about $6.5\ GHz/12.5\ GHz = 0.52$. The residual small deviation might be attributed to the measurement error of component losses.

| Components | Losses (dB) |
|---|---|
| PPLN waveguide output coupling | 2.50 |
| DWDM-Signal | 1.83 |
| DWDM-Heralding | 2.21 |
| Narrowband DWDM-$f_{i1}$ | 4.58 |
| Narrowband DWDM-$f_{i2}$ | 4.58 |
| Narrowband DWDM-$f_{i3}$ | 4.60 |
| Delay fiber | 0.40 |
| EOM | 4.56 |
| TNF | 5.30 |
| SNSPD | 1.80~2.20 |

**Table S1.** Overall component losses of the signal and heralding arms. PPLN: Periodically-Poled Lithium Niobate; DWDM: Dense Wavelength Divided Multiplexer; EOM: Electro-Optic phase Modulator; TNF: Tunable Narrowband Filter; SNSPD: Superconducting Nanowire Single Photon detector.

| Efficiencies | % |
|---|---|
| Transmission efficiency-Signal arm | 2.10 |
| Transmission efficiency-Heralding arm | 7.06 |
| Collection efficiency-Signal arm | 2.15 |
| Collection efficiency- $f_{i1}$ | 5.34 |
| Collection efficiency- $f_{i2}$ | 4.87 |
| Collection efficiency- $f_{i3}$ | 5.43 |

**Table S2.** Total transmission efficiencies and collection efficiencies of the signal and heralding arms.

**References**

1. Ou, Z. Quantum theory of fourth-order interference. *Phys. Rev. A* **37**, 1607 (1988).

2. Li, X., Yang, L., Cui, L., Ou, Z. Y. & Yu, D. Observation of quantum interference between a single-photon state and a thermal state generated in optical fibers. *Opt. Exp.* **16**, 12505–12510 (2008).

3. Sun, F., Wong, C. et al. Indistinguishability of independent single photons. *Phys. Rev. A* **79**, 013824 (2009).

4. Avenhaus, M. et al. Photon number statistics of multimode parametric down-conversion. *Phys. Rev. Lett.* **101**, 053601 (2008).